\documentclass[aps,prb,twocolumn]{revtex4} 

\usepackage{graphicx}
\usepackage{dcolumn}
\usepackage{bm}

\newcommand{\comment}[1]{}

\begin{document}
\renewcommand{\theequation}{\arabic{section}.\arabic{equation}}

\title{Fermionic Phonons:
Exact Analytic Results
and Quantum Statistical Mechanics
for a One Dimensional Harmonic Crystal.}


\author{Phil Attard}
\affiliation{{\tt phil.attard1@gmail.com}}


\begin{abstract}
Analytic expressions for the energy eigenvalues and eigenfunctions
of a one-dimensional harmonic crystal are obtained.
The average energy and density profiles
are obtained numerically as a function of temperature.
A surprisingly large number of  energy levels
(eg.\ 5,000 for 4 particles)
are required for reliable results
at even moderate temperatures.
\end{abstract}

\pacs{}

\maketitle

%
\section{Introduction}
\setcounter{equation}{0} \setcounter{subsubsection}{0}
%


Exact analytic results for model systems have proved useful
in the development of the physical sciences.
They provide benchmarks against which to test approximate techniques,
and they give insight into the mechanisms
for the physical behavior of more realistic systems.
Although generally those models that are amenable to exact analytic solution
are necessarily a simplification of reality,
the results do have the  great advantage of being unambiguous
and free from doubts concerning approximations, convergence,
numerical techniques etc.
Further, the parameter space may be rapidly explored,
allowing general conclusions to be drawn,
and numerically sensitive regimes to be identified.

This paper treats a model of a one-dimensional crystal
with nearest neighbor harmonic interactions.
The potential energy is a quadratic form,
with energy eigenvalues and eigenvectors being obtained explicitly.
These allow the vibrational modes to be identified,
and the exact quantum mechanical solution to be invoked.

The model is realistic in that it includes particle interactions.
It is therefore more sophisticated
than the quantum ideal gas,
or the independent quantum harmonic oscillator
that are the routine examples studied by quantum statistical mechanics.
Because of the interactions,
new insight is provided into phenomena such
as collective motion, vibrational excitations, and  phonons
that is not available with the ideal systems.

An example of the utility of the present exact model calculations
is that even for a small  system (eg.\ 4 or 5 particles),
5,000 energy levels are  required to obtain
quantitatively accurate results at the moderately high temperatures
where classical effects become noticeable,
which is typical for terrestrial condensed matter.
The computational advantage of the present analytic model
can be quantified by comparison with the work of
Hernando and Van\'i\v cek,\cite{Hernando13}
who, for 4 or 5 interacting Lennard-Jones particles,
obtained numerically 50 energy eigenvalues.

%
\section{Exact Analysis}
\setcounter{equation}{0} \setcounter{subsubsection}{0}
%

\subsection{Model}

Following earlier work by the author,\cite{TDSM}
consider a one-dimensional harmonic crystal in which the particles
are attached by linear springs to each other and to lattice sites.
Let the coordinate of the $j$th particle
be $q_{j}$, and let its lattice position
(ie.\ in the lowest energy state) be $\overline q_{j} = j \Delta_q$.
The lattice spacing is also the relaxed inter-particle spring length.
There are fixed `wall' particles at $q_{0} = 0$
and $q_{N+1} = (N+1)\Delta_q$.
Let $d_j \equiv q_{j}- \overline q_{j}$
be the displacement from the lattice position;
for the wall particles, $d_0=d_{N+1}=0$.
The system has over-all number density $\rho = \Delta_q^{-1}$.

In this model, there is an external harmonic potential
of spring constant $\kappa$
acting on each particle centered at its lattice site.
The inter-particle spring has strength $\lambda$
and relaxed length $\Delta_q$.
With these the potential energy is
\begin{eqnarray}  \label{Eq:U(q)}
U({\bf q})
& = &
\frac{\kappa}{2} \sum_{j=1}^N [q_{j}-\overline q_{j}]^2
+
\frac{\lambda}{2} \sum_{j=0}^N [q_{j+1}-q_{j}-\Delta_q]^2
\nonumber \\ & = &
\frac{\kappa}{2} \sum_{j=1}^N d_j^2
+
\frac{\lambda}{2} \sum_{j=0}^N [d_{j+1}-d_j]^2
\nonumber \\ & = &
\frac{-\lambda}{2} \underline{ \underline K}^{(N)}:{\bf d}{\bf d}.
\end{eqnarray}
Here $\underline{ \underline K}^{(N)}$ is an $N \times N$ tridiagonal matrix
with elements
\begin{equation}
K_{jk} =
\left\{ \begin{array}{ll}
K , & j=k \\
1 , & j = k \pm 1 , \\
0 , & \mbox{ otherwise,}
\end{array} \right.
\end{equation}
where $K \equiv -2 - \kappa/\lambda$.

It should be mentioned explicitly
that the lattice positions are in order,
$\overline q_j < \overline q_{j+1}$, $j=1,2,\ldots,N$.
However for the particle positions themselves
there is no similar constraint on their order.

It is axiomatic in quantum mechanics
that the Hamiltonian operator must be fully symmetric
with respect to the permutation of identical particles.
\cite{Messiah61,Merzbacher70}
Otherwise the symmetry of the wave function
would not be preserved during its evolution,
which is to say that bosons would decay into fermions,
and \emph{vice versa}.
The above potential energy is not symmetric
(eg.\ in transposing particle positions
$q_{j}$ and $q_{k}$,
one replaces the respective one-body terms with
$\kappa [q_{k}-\overline q_{j}]^2 /2$ and
$\kappa [q_{j}-\overline q_{k}]^2 /2$,
as well as the respective pair interactions with
$ \lambda [q_{j \pm 1}-q_{k} \mp \Delta_q]^2/2$
and
$ \lambda  [q_{k \pm 1}-q_{j} \mp \Delta_q]^2/2$,
which changes the value of the potential energy).
This means that the particles are not identical
and that the wave function is not restricted by symmetrization.
The original version of this paper
was vitiated by the author's misunderstanding of this point.

\subsection{Eigenvalues and Eigenvectors}

The eigenvalues of the potential energy matrix are required,
and these may be obtained from the characteristic equation,
\begin{eqnarray}
S_N(K-\mu)
& \equiv &
\left| \underline{ \underline K}^{(N)} - \mu {\mathrm I} \right|
\rule[-.5cm]{0cm}{1cm}
 \\  & = &
\left| \begin{array}{ccccc}
K - \mu  & 1 & 0 & \ldots & 0 \\
1 & K - \mu  & 1 & 0 & \ldots   \\
0 & 1 & K - \mu  & 1 & \ldots   \\
\vdots & && \ddots & \vdots \\
0 & \ldots & 0 & 1 & K - \mu
\end{array}  \right|
\nonumber \\ & = &
(K - \mu)
\left| \underline{ \underline K}^{(N-1)} - \mu {\mathrm I} \right|
\nonumber \\  & &
-
\left| \begin{array}{ccccc}
1  & 1 & 0 & \ldots & 0 \\
0 & K - \mu  & 1 & 0 & \ldots   \\
0 & 1 & K - \mu  & 1 & \ldots   \\
\vdots & && \ddots & \vdots \\
0 & \ldots & 0 & 1 & K - \mu
\end{array}  \right|
\nonumber \\ & = &
(K - \mu) S_{N-1}(K-\mu)
- S_{N-2}(K-\mu) .\nonumber
\end{eqnarray}
This is just the recursion relation
for the Tchebyshev polynomials of the second kind,
which are denoted $S_n(x) = U_n(x/2)$
by Abramowitz and Stegun, Eq.~(AS22.7.6).\cite{Abramowitz70}
They give $S_N(2 \cos \theta) = U_N(\cos \theta) =
\sin((N+1)\theta)/\sin \theta$,
which evidently vanishes when $\theta_n = \pm n\pi/(N+1)$,
$n=1,2,\ldots,N$.
Hence the characteristic equation vanishes when
$K-\mu_n = 2 \cos \theta_n $, or
\begin{equation}
\mu_n = K + 2 \cos \frac{n\pi}{N+1}
, \;\; n=1,2, \ldots , N.
\end{equation}
These are the eigenvalues,
and since $K \equiv -2 - \kappa/\lambda$,
they are negative.

The corresponding eigenvectors, ${\bf u}_n$, have elements
\begin{equation}
u_{n,j} =
\sqrt{\frac{2}{N+1}} \sin \frac{jn\pi}{N+1} ,\;\;
j=1,2,\ldots, N.
\end{equation}
It may be shown that these form an orthonormal set.

\subsection{Normal Modes}

The matrix of eigenvectors,
\begin{equation}
\underline{ \underline X}
\equiv
\{ {\bf u}_1,{\bf u}_2, \ldots , {\bf u}_N\}
=
\left( \begin{array}{cccc}
u_{1,1}  & u_{2,1} & \ldots & u_{N,1} \\
u_{1,2}  & u_{2,2} & \ldots & u_{N,2} \\
\vdots & &\ddots & \vdots \\
u_{1,N}  & u_{2,N} & \ldots & u_{N,N}
\end{array}  \right) ,
\end{equation}
is  orthogonal,
$ \underline{ \underline X}^\mathrm{T} \underline{ \underline X}
= \underline{ \underline X}\, \underline{ \underline X}^\mathrm{T}
=\mathrm{I}$.
(It is also symmetric,
$ \underline{ \underline X}^\mathrm{T} = \underline{ \underline X}$.)
With this the potential energy may be written
\begin{eqnarray}
U({\bf q})
& = &
\frac{-\lambda}{2}
\underline{ \underline K} : {\bf d} {\bf d}
\nonumber \\ & = &
\frac{-\lambda}{2}
( \underline{ \underline X}^\mathrm{T}  {\bf d} )^\mathrm{T}
\underline{ \underline X}^\mathrm{T}
\,
\underline{ \underline K}
\,
\underline{ \underline X}
( \underline{ \underline X}^\mathrm{T}  {\bf d} )
\nonumber \\ & = &
\frac{-\lambda}{2}
{\bf q}'^{\mathrm T} \underline{ \underline D} {\bf q}'.
\end{eqnarray}
Here $\underline{ \underline D} \equiv
\underline{ \underline X}^\mathrm{T}  \underline{ \underline K}
\, \underline{ \underline X}$
is diagonal,
$D_{n,n'} = \mu_n \delta_{n,n'}$,
and the modes are defined as
\begin{equation}
{\bf q}'
\equiv
\underline{ \underline X}^\mathrm{T} {\bf d} .
\end{equation}

Suppose that the particles have mass $m$,
and that the momentum of the $j$th particle is $p_{j} = m \dot q_j$.
Hence the classical kinetic energy is
\begin{equation}
{\cal K}({\bf p})
= \frac{1}{2m}  {\bf p} \cdot {\bf p}
= \frac{1}{2m}  {\bf p}' \cdot {\bf p}' ,
\end{equation}
where ${\bf p}' \equiv \underline{ \underline X}^\mathrm{T} {\bf p}$.
Accordingly the classical Hamiltonian is
\begin{equation}
{\cal H}({\bf \Gamma})
= {\cal K}({\bf p}) + U({\bf q})
=
\frac{1}{2m}  {\bf p}' \cdot {\bf p}'
-\frac{\lambda}{2}
\underline{ \underline D} : {\bf q}'{\bf q}' .
\end{equation}
Evidently, the normal modes represent independent harmonic oscillators.

\subsection{Quantum Mechanics}

The Hamiltonian operator
in the normal mode representation is
\begin{eqnarray}
\hat{\cal H}
& = &
\frac{1}{2m} \hat {\bf p}' \cdot \hat {\bf p}'
-\frac{\lambda}{2}
\underline{ \underline D} : {\bf q}'{\bf q}'
\nonumber \\ & = &
\frac{1}{2m}  \sum_{n=1}^N  \hat p_n'^2
-\frac{\lambda}{2}
\sum_{n=1}^N \mu_n q_n'^2
\nonumber \\ & = &
\sum_{n=1}^N \frac{\hbar \omega_n}{2} \left\{ \hat P_n^2 + Q_n^2 \right\} ,
\end{eqnarray}
where $ \hat P_n \equiv \hat p_n'/\sqrt{m \hbar \omega_n}$,
 $ Q_n \equiv q_n' \sqrt{m  \omega_n/\hbar}$,
and $ m \omega_n^2 \equiv - \lambda \mu_n$.
Since $\mu_n < 0$ and $\lambda > 0$,
the frequencies are real.

For each mode, the energy eigenvalues are\cite{Messiah61,Merzbacher70}
\begin{equation}
E_{n,l_n} = (l_n + 0.5) \hbar \omega_n ,
\;\; l_n = 0,1,2, \ldots ,
\end{equation}
and the corresponding energy eigenfunctions
are the Hermite functions,\cite{Messiah61,Merzbacher70,WikiQHO}
\begin{equation}
\phi_{n,l_n}(Q_n) \equiv
\frac{1}{\sqrt{2^{l_n} {l_n}! \sqrt{\pi}}}
e^{-Q_{n}^2/2}
\mathrm{H}_{l_n}(Q_n) ,
\end{equation}
where $\mathrm{H}_{l_n}(Q)$ is the Hermite polynomial of degree $l_n$.
The eigenfunctions of the system have product form,
\begin{equation}
\phi_{\bf l}({\bf Q})
= \prod_{n=1}^N \phi_{n,l_n}(Q_n) ,
\end{equation}
and the energy in such an eigenstate is of course
$E_{\bf l} = \sum_{n=1}^N E_{n,l_n}$.
Since the mode amplitudes are a function of the positions,
the energy eigenfunction will often instead be written
$\phi_{\bf l}({\bf q})$.

%
\section{Results}
\setcounter{equation}{0} \setcounter{subsubsection}{0}
%

\subsection{Computational Details}

The computational implementation of the eigenvalue and eigenvector
calculations is obvious and need not be detailed here.
There is perhaps one interesting algorithmic challenge
for the quantum statistical mechanical aspects of the problem.

Because of the Maxwell-Boltzmann factor,
it was  useful
to order the possible energy states beginning with the lowest.
The $N$ mode frequencies $\omega_n$ are known,
and one has to order the possible sets of quanta,
${\bf l} = \{ l_1,l_2,\ldots,l_N\}$,
in terms of their energy,
$E_{{\bf l}(l)} \le E_{{\bf l}(l+1)}$, $l=1,2,\ldots$.
This was done by initially creating $l^\mathrm{max}$ states
from the lowest frequency mode only,
${\bf l}(l) = \{ l,0,0,\ldots,0\}$, $l=1,2,\ldots,l^\mathrm{max}$.
Then, one quantum of the next higher frequency mode was added
and the corresponding state inserted in order,
bumping up the higher energy states,
and discarding the highest.
This continued until the insertion point reached $l^\mathrm{max}$,
at which point the cycle was repeated for the next higher frequency.

\subsection{Units}

In order to make contact with a real physical system,
in the results below parameters for a Lennard-Jones model of
neon may be used.
These are a mass $m=3.35 \times 10^{-26}$\,kg,
a separation of zero force $r_\mathrm{e} = 3.13 \times 10^{-10}$\,m,
and a Lennard-Jones well-depth $\varepsilon = 4.93 \times 10^{-22}$\,J.
\cite{Sciver12}

The Lennard-Jones pair potential,
$u_\mathrm{LJ}(r)
=\varepsilon [ (r_\mathrm{e} /r)^{12} - 2(r_\mathrm{e} /r)^{6} ]$,
allows one to define the Lennard-Jones  frequency,
$\omega_\mathrm{LJ} = \sqrt{ 72 \varepsilon/(m r_\mathrm{e}^2 )}$,
which equals $3.29 \times 10^{12}$\,Hz for neon.
This is derived from the curvature of the Lennard-Jones potential
at the zero force separation.
The results below are mainly expressed in units of
the Lennard-Jones zero force separation $r_\mathrm{e} $,
and the Lennard-Jones frequency $\omega_\mathrm{LJ}$.

The thermal wave length,
$\Lambda_\mathrm{th} \equiv \sqrt{ 2\pi\hbar^2/(m k_\mathrm{B}T) }$,
can be used as a guide to the importance of symmetrization effects.
(Here $\hbar$ is Planck's constant,
$k_\mathrm{B}$ is Boltzmann's constant,
and the convenient inverse temperature is
$\beta \equiv 1/k_\mathrm{B}T$.)
When the thermal wave length is comparable to, or larger than,
the lattice spacing
$\Delta_q \equiv \overline q_{j+1} - \overline q_{j} $,
symmetrization effects should be measurable.
They are also measurable when the spring constants are small.

\subsection{Numerical Results}

\begin{figure}[t!]
\centerline{
\resizebox{8cm}{!}{ \includegraphics*{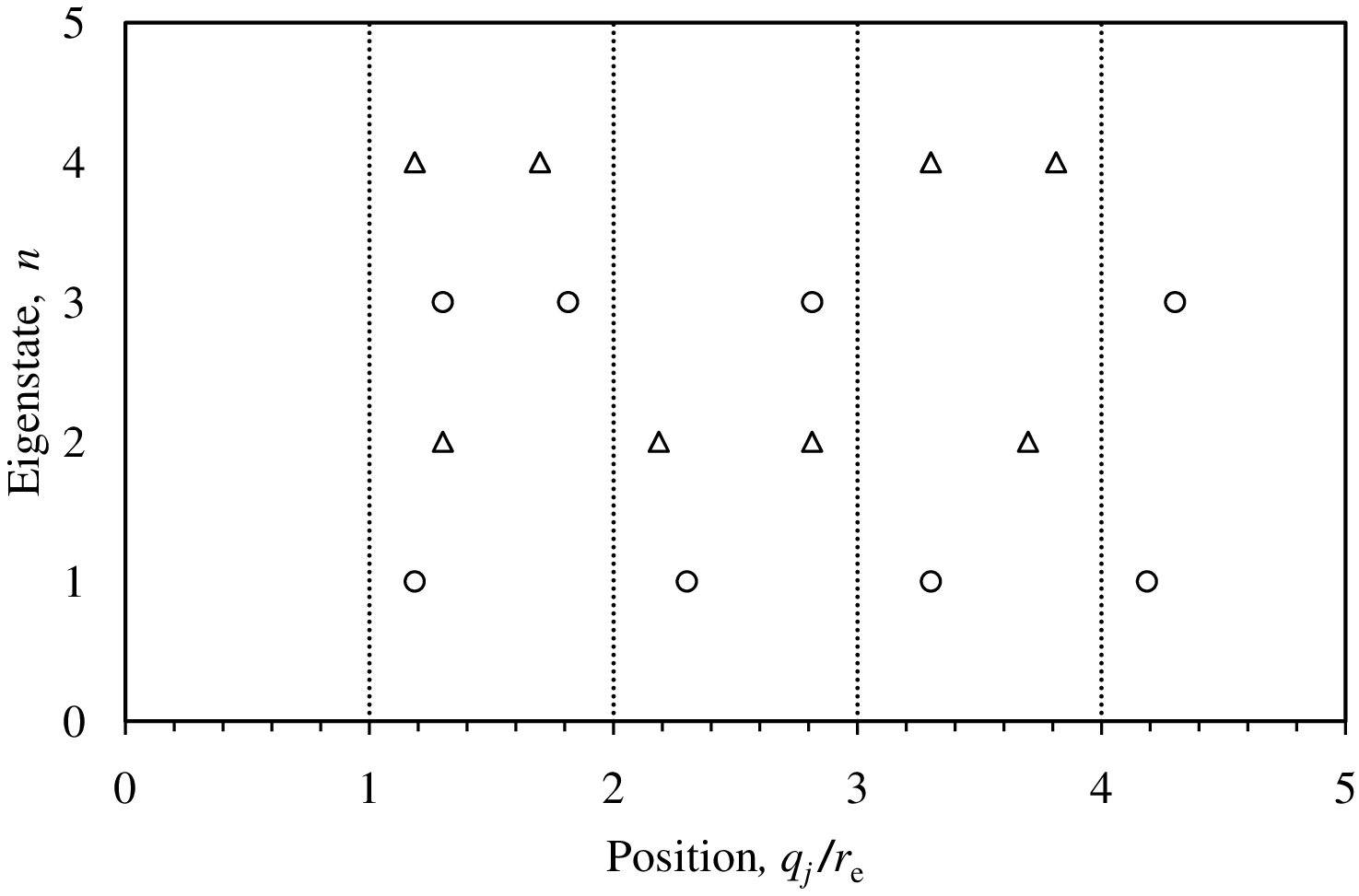} } }
\caption{\label{Fig:q0j(un)}
Eigenvectors of the potential energy matrix,
as described by positions of the  $N=4$ particles
($\lambda = \kappa = m\omega_\mathrm{LJ}^2$).
The corresponding eigenvalues increase in magnitude
from bottom to top.
The dotted lines indicate the lattice positions $\overline q_{j}$.
The displacements are all scaled by the same amount.
}
\end{figure}

Figure \ref{Fig:q0j(un)} shows the positions of the particles
in the eigenvectors of the potential energy matrix
for the four particle harmonic crystal.
For the lowest frequency mode,
the particles are all displaced in the same direction (in phase),
with the central two particles having twice the amplitude
of the outer two.
For the second lowest frequency,
the middle pair of particles are out of phase.
For the second highest frequency,
both outer pairs of particles are out of phase.
And finally, for the highest frequency,
all three consecutive pairs of particles are out of phase.

\begin{figure}[t!]
\centerline{
\resizebox{8cm}{!}{ \includegraphics*{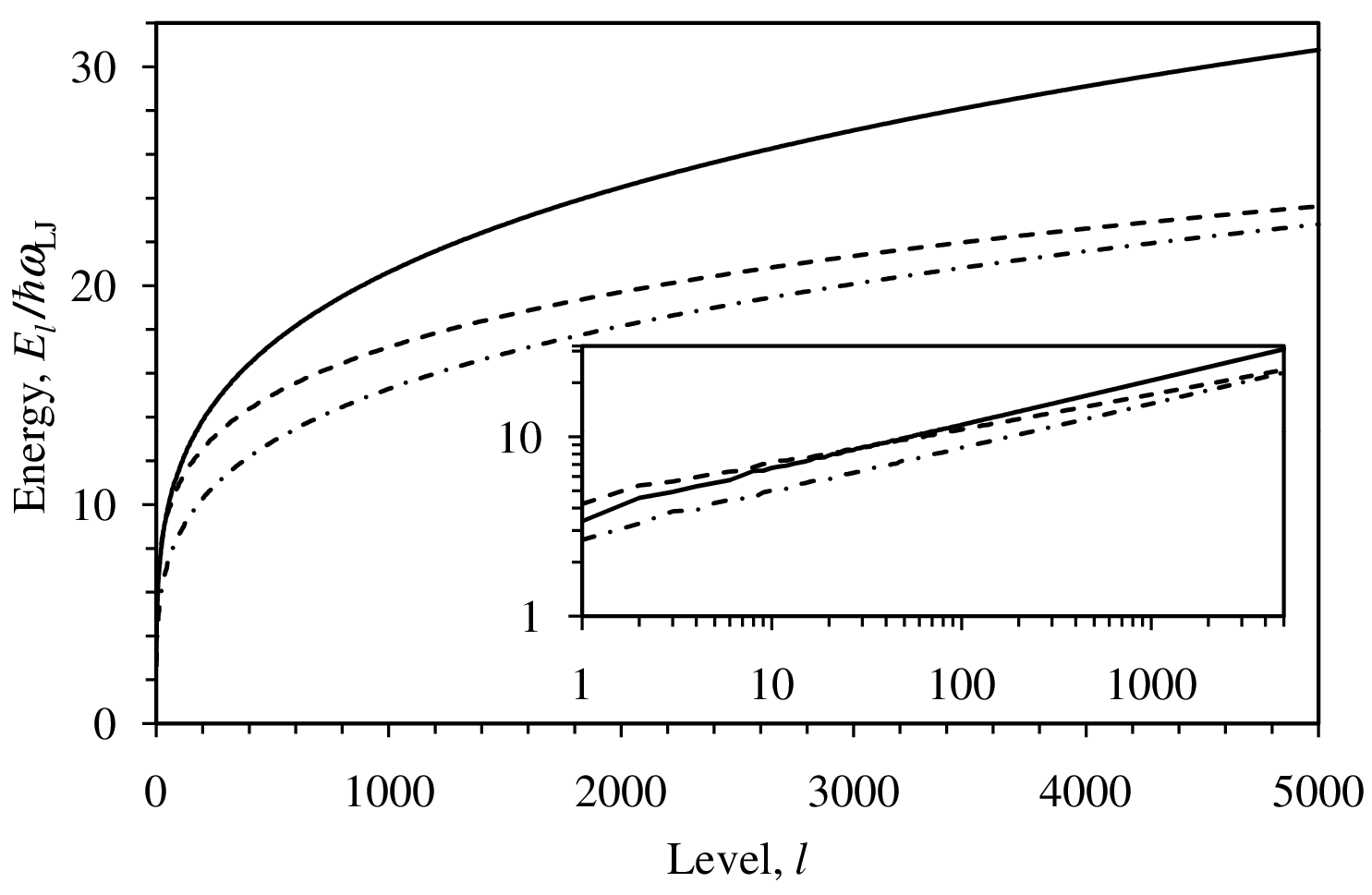} } }
\caption{\label{Fig:El}
The first 5,000 quantum energy levels
for $N=4$, $\lambda = \kappa = m\omega_\mathrm{LJ}^2$
(solid curve);
for $N=4$, $\lambda  = m\omega_\mathrm{LJ}^2$, $\kappa=0$,
(dash-dotted curve);
and for $N=5$, $\lambda = \kappa = m\omega_\mathrm{LJ}^2$
(dashed curve).
In all cases, $\Delta_q = r_\mathrm{e}$.
{\bf Inset.} Log-log plot.
}
\end{figure}

Figure \ref{Fig:El}
shows the quantized energy levels for the harmonic crystal
for several sets of parameters.
Initially the energy of each level increases rapidly
with level number.
But for large level numbers,
the rate of change of energy with energy level is sub-linear.
Or to put it another way,
the density of energy states increases with increasing energy.
For $l \agt 10$, the data is well fitted by $E_l \propto l^{1/N}$,
as can be seen in the inset to the figure.

\begin{figure}[t!]
\centerline{
\resizebox{8cm}{!}{ \includegraphics*{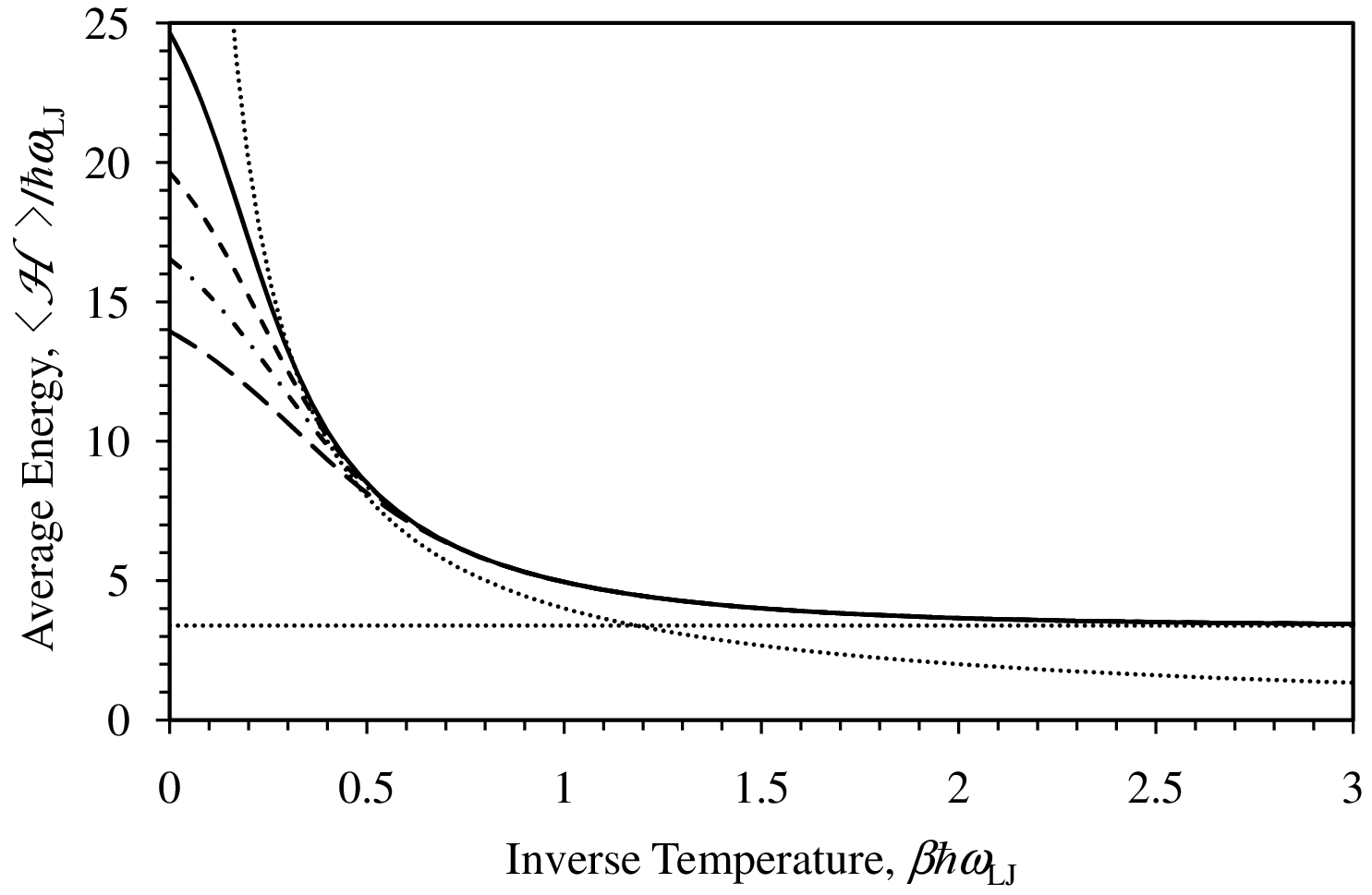} } }
\caption{\label{Fig:EvsB}
Average energy calculated with various numbers of levels
($N=4$, $\Delta_q = r_\mathrm{e}$,
$\lambda = \kappa = m\omega_\mathrm{LJ}^2$
and $d=0$).
From bottom to top,
$l_\mathrm{max} =$ 500 (long dashed), 1,000 (dash-dotted),
2,000 (short dashed), and 5,000 (full).
The dotted curve is the classical result,
$\langle E \rangle_\mathrm{cl}=N/\beta$.
The dotted line is the energy of the ground state,
$E_1 = \sum_{n=1}^N \hbar \omega_n/2$.
}
\end{figure}

Figure \ref{Fig:EvsB} shows the average energy
for a canonical equilibrium system
as a function of inverse temperature.
The data tests the dependence of the average on the number of levels used,
$l_\mathrm{max}$.
One can see that for $\beta \hbar \omega_\mathrm{LJ} \agt 0.8$,
the results for all $l_\mathrm{max} \ge 500$ are indistinguishable.
For $\beta \hbar \omega_\mathrm{LJ} \alt 0.5$,
there is a discernable difference between
$l_\mathrm{max} =$ 2,000 and 5,000,
and this difference increases with  increasing temperature
(decreasing inverse temperature).

The figure also shows the energy of the ground state,
$E_1 = \sum_{n=1}^N \hbar \omega_n /2$.
It can be seen that for these parameters,
excited states make negligible contribution
when the temperature is lower than
$\beta \hbar \omega_\mathrm{LJ} \agt 2$.

The present  crystal has $2N$ harmonic modes in the classical Hamiltonian
($N$ in the potential energy, and $N$ in the kinetic energy).
Hence by the equipartition theorem,
the average energy is $\langle {\cal H} \rangle_\mathrm{cl}
= N/\beta $,
which is the dotted curve in Fig.~\ref{Fig:EvsB}.
One can see that this lies increasingly below the quantum results
as the temperature  decreases (inverse temperature increases),
and it lies increasingly above the quantum results
as the temperature increases.
In the present case there is a region,
$ 0.3 \alt \beta \hbar \omega_\mathrm{LJ}  \alt 0.4$,
in which the quantum results for $l_\mathrm{max} =  5,000$
coincide with the classical result.
The data suggests what one knows to be true:
the classical result must be the limiting result at high temperatures,
but an increasing number of energy levels
contribute to the average as the temperature increases.
Hence for fixed  $l_\mathrm{max}$,
there is always a temperature above which the quantum results
become inaccurate.

One can draw the important conclusion from this figure
that starting from the exact quantum approach
is a very inefficient way of obtain the classical result,
which is the exact high temperature limit.

\begin{figure}[t!]
\centerline{
\resizebox{8cm}{!}{ \includegraphics*{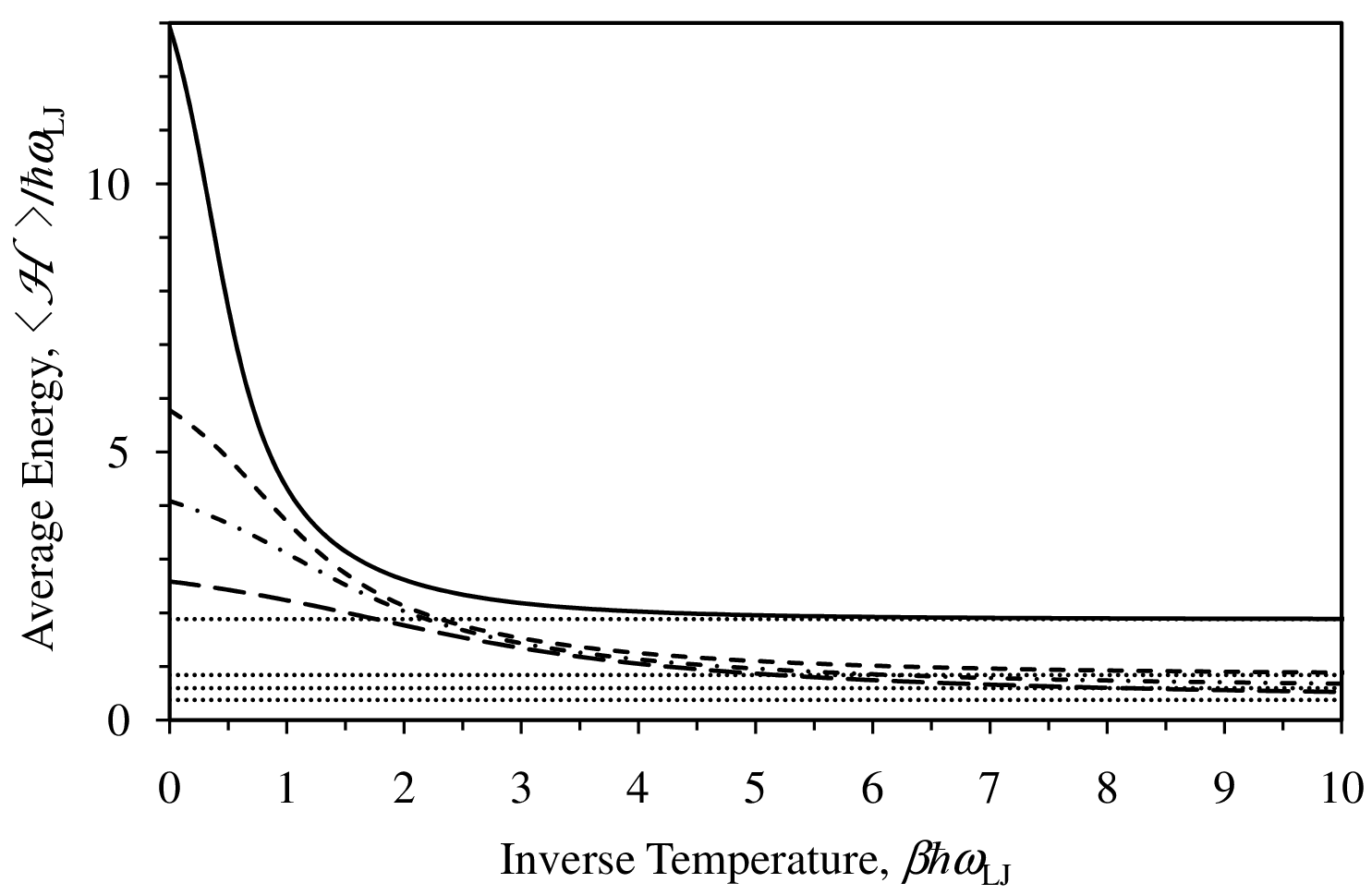} } }
\caption{\label{Fig:EvsB2}
Average energy for various values
of the interparticle spring constant $\lambda$
($N=4$, $\Delta_q = r_\mathrm{e}$,
$ \kappa = 0 $,
$l_\mathrm{max} =5,000$, and $d_{\mathrm m}=0$).
From top to bottom,
$\lambda/ m\omega_\mathrm{LJ}^2 =$
0.5 (solid curve),
0.1 (short dashed curve),
0.05 (dash-dotted curve,
and 0.02 (long dashed curve).
The dotted lines are the respective ground states.
}
\end{figure}

Figure \ref{Fig:EvsB2} shows the average energy
for several values of the interparticle spring constant $\lambda$.
In the  low temperature limit, $\beta \rightarrow \infty$,
the average energy is the ground state energy,
$E_0 = \sum_{n=1}^N \hbar \omega_n /2$.
It can be seen in Fig.~\ref{Fig:EvsB2} that for
$\beta \hbar \omega_\mathrm{LJ} \agt $ 3--9
($\lambda / m\omega_\mathrm{LJ}^2 =$ 0.5--0.02)
the system may be considered to be in the ground state.

\begin{figure}[t!]
\centerline{
\resizebox{8cm}{!}{ \includegraphics*{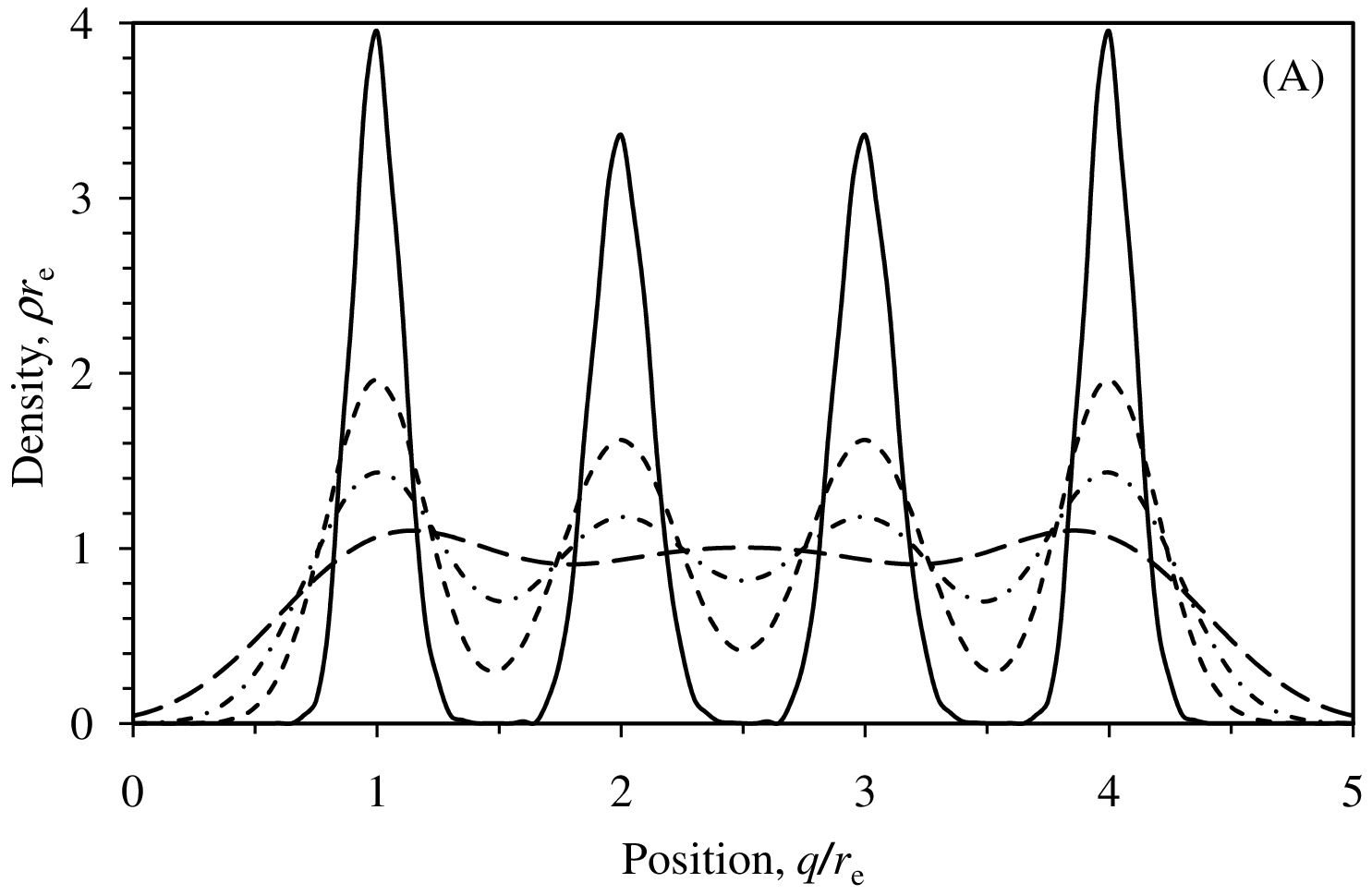} } }
\caption{\label{Fig:rhovsq0}
Density profiles for various values
of the interparticle spring constant $\lambda$
($N=4$, $\Delta_q = r_\mathrm{e}$,
$ \kappa = 0 $,
$l_\mathrm{max} =5,000$,
and $\beta \hbar \omega_\mathrm{LJ} = 2$).
From top to bottom at the peaks,
$\lambda/ m\omega_\mathrm{LJ}^2 =$
0.5 (solid curve),
0.1 (short dashed curve),
0.05 (dash-dotted curve),
and 0.02 (long dashed curve).
}
\end{figure}

Figure \ref{Fig:rhovsq0} shows the density profiles
corresponding to the same cases as the preceding figure
at the temperature $\beta \hbar \omega_\mathrm{LJ} =2$.
The profiles are normalized to integrate to $N=4$.
In general the peaks and troughs in the profiles
indicate that the particles are
mainly localized to their respective lattice positions.
At the highest interparticle spring constant shown,
the density is zero between lattice points,
which means that there is little overlap between the particles.
Conversely, at the lowest coupling shown,
the density peaks are much broader
and there is a high probability of finding a particle
between the lattice positions.

%
\section{Conclusion}
\setcounter{equation}{0} \setcounter{subsubsection}{0}
%

This paper has obtained analytic expressions for the energy eigenvalues
and eigenfunctions of a realistic system composed of interacting particles,
namely a one dimensional harmonic crystal.
Although the results are mainly practical and utilitarian in nature,
there is one generic conceptual point that emerges:
because wave function symmetrization
does not apply to crystals (i.e.\ solids in which the particles
may be distinguished by their attachment to fixed points in space),
the consequent quantized vibrational modes,
otherwise known as phonons,
are not bosons.
It would be more accurate to call them
`non-bosonic phonons' rather than  `fermionic phonons',
as in the title to this paper.
This is a general conceptual point that holds beyond the one-dimensional
harmonic crystal treated here.

As mentioned the bulk of the paper is concerned with
numerical results for the  one-dimensional harmonic crystal.
A number of quantitative conclusions can be drawn from these,
such as the degree to which the average energy increases,
and the structure in the density profile decreases,
with increasing temperature.

The advantage of being able to easily explore the parameter space
with the analytic model
can be illustrated by comparison with the exact but numerical studies
of Hernando and Van\'i\v cek,\cite{Hernando13}
who, with praiseworthy effort,
obtained the first 50 energy eigenvalues
of a one-dimensional Lennard-Jones system.
Of course their Lennard-Jones particles
confined by a weak harmonic potential
is not the same as the present particles interacting with linear springs
and confined by fixed wall particles.
Nevertheless, in so far as the separation between the Lennard-Jones
particles is approximately $r_\mathrm{e}$,\cite{Hernando13}
the Lennard-Jones frequency
$\omega_\mathrm{LJ} = \sqrt{ 72 \varepsilon/(m r_\mathrm{e}^2 )}$
seems the appropriate energy scale
to use in qualitatively comparing the two systems.

One point of interest is that
Hernando and Van\'i\v cek\cite{Hernando13}
obtained the average density profile for 4 particles
for several temperatures,
namely $\beta \hbar \omega_\mathrm{LJ} = $ 6.8, 11.9, and 47.5
in the present units.
They found marked changes in the profiles,
with the highest temperature case losing almost completely the density peaks
present at the lower temperatures.
This means that the excited states contribute significantly
to their higher temperature results.
The fact that  Fig.~\ref{Fig:EvsB}
for the present harmonic crystal at $\rho = r_\mathrm{e}^{-1}$
shows no evidence of excited states for
$\beta \hbar \omega_\mathrm{LJ} \agt 2$,
and significant change from the ground state
only for $\beta \hbar \omega_\mathrm{LJ} \alt 0.5$,
suggests that the inverse temperature for the harmonic crystal
should be multiplied by a factor of 10--20 for the purposes of comparison
with the Lennard-Jones system.
Given that the 50 excited states
obtained by Hernando and Van\'i\v cek\cite{Hernando13}
make a significant difference to their density profiles,
it is questionable whether or not 50 energy eigenvalues are actually enough
to accurately describe the system at those particular temperatures.
In particular, the present Fig.~\ref{Fig:EvsB} shows
that 5,000 energy eigenvalues are necessary for the harmonic crystal
for $\beta \hbar \omega_\mathrm{LJ} \alt 0.5$,
which, if scaled by 10--20,
would be comparable to the higher temperatures used in the earlier study.
Comparing the two different systems is obviously of questionable validity,
but it is intriguing that
the  mean field classical phase space
calculations of the present author\cite{Attard18c}
were in better agreement with the lower temperature results
of Hernando and Van\'i\v cek\cite{Hernando13}
than with their higher temperature ones.

In any case the larger lesson from the present exact calculations
is that in the terrestrial regime,
where quantum effects are comparable to,
or a perturbation on the classical result,
it is necessary to obtain a prohibitively large
number of energy eigenvalues for even quite a small system.
One can conclude from this
that it is better to treat terrestrial condensed matter systems
as a quantum perturbation of a classical system,
rather than as a fully quantum system.
It is hoped quantitatively to  confirm or refute this conclusion
with explicit classical phase space calculations
for the present model in a future publication.



\appendix

%
\section{Symmetrization for Spin-Position Factorization} \label{Sec:spin}
\setcounter{equation}{0} \setcounter{subsubsection}{0}
\renewcommand{\theequation}{\Alph{section}.\arabic{equation}}
%

This appendix is independent of the one-dimensional harmonic crystal
explored in the text.

The set of commuting dynamical variables for one particle $j$
may be taken to be
${\bf x}_j = \{ {\bf q}_j,\sigma_j\}$,
where $\sigma_j \in \{-S,-S+1,\ldots,S\}$
is the $z$-component of the spin of particle $j$.
(See Messiah (1961) \S 14.1 or Merzbacher (1970) \S 20.5.)
\cite{Messiah61,Merzbacher70}
Note that here $\sigma$
is \emph{not} a spin operator or a Pauli spin matrix.
Label the $2S+1$ spin eigenstates of particle $j$ by
$s_j \in \{-S,-S+1,\ldots,S\}$,
and the spin basis function by $\alpha_{s_j}(\sigma_j) =
\delta_{s_j,\sigma_j}$.
Note that this is \emph{not} a spinor.
For $N$ particles,
${\bm \sigma} \equiv \{ \sigma_1, \sigma_2, \ldots, \sigma_N \}$,
and similarly for ${\bf s}$ and ${\bf q}$,
and the basis functions for spin space are
$\alpha_{\bf s}({\bm \sigma}) =
\delta_{{\bf s},{\bm \sigma}}
= \prod_{j=1}^N \delta_{s_j,\sigma_j}$.
The states ${\bf n}$ may be single or multi-particle states.

An unsymmetrized wave function $ \psi({\bf x})$
in general has symmetrized form
\begin{equation}
\psi^\pm({\bf x})
\equiv
\frac{1}{\sqrt{\chi^\pm N!}}
\sum_{\hat{\mathrm P}} (\pm 1)^{p}
\psi(\hat{\mathrm P}{\bf x}),
\end{equation}
with the symmetrization factor being
\begin{equation}
\chi^\pm \equiv
\sum_{\hat{\mathrm P}} (\pm 1)^{p}
\langle  \psi(\hat{\mathrm P}{\bf q}) | \psi({\bf q}) \rangle .
\end{equation}

Alternatively,
one can expand the wave function in terms of spin-position basis functions,
\begin{equation}
\psi({\bf x})
= \sum_{{\bf s},{\bf n}}
\langle \alpha_{\bf s}\phi_{\bf n}| \psi \rangle \,
\Phi_{{\bf n},{\bf s}}({\bf x})
, \;\;
\Phi_{{\bf n},{\bf s}}({\bf x})
\equiv
\alpha_{\bf s}({\bm \sigma})\phi_{\bf n}({\bf q}) .
\end{equation}
Here and below the $\alpha_{\bf s}$ will be called the spin basis functions,
and the $\phi_{\bf n}$ will be called the position basis functions.
This nomenclature favors brevity over precision;
better might be, for example,
the basis functions for spin and position space, respectively.
Instead of position one could use the momentum representation.
It will often prove useful to choose  the $\phi_{\bf n}$
to be energy eigenfunctions.
The $\alpha_{{\bf s}}({\bm \sigma})$ and the $\phi_{{\bf n}}({\bf q})$
form a complete orthonormal set.

The symmetrization of any wave function in  spin-position space
can be accomplished by using symmetrized basis functions
in its expansion.
The latter are given by
\begin{equation} \label{Eq:Phi-exact}
\Phi^\pm_{{\bf n},{\bf s}}({\bf x})
=
\frac{1}{\sqrt{\chi^\pm_{{\bf n},{\bf s}} N!}}
\sum_{\hat{\mathrm P}}
(\pm 1)^{p}
\alpha_{\bf s}(\hat{\mathrm P}{\bm \sigma})
\phi_{\bf n}(\hat{\mathrm P}{\bf q}) .
\end{equation}
This is exact.

In place of this exact symmetrization,
one can invoke an approximation that relies upon
the factorization of the spin-position  basis function
into the sum of products of symmetrized position basis functions
and  symmetrized spin basis functions,
namely
\begin{equation} \label{Eq:Phi-approx}
\Phi_{{\bf n},{\bf s}}^\pm({\bf x})
=
\left\{ \begin{array}{l}
\displaystyle
\frac{1}{\sqrt{\tilde \chi^+_{{\bf n},{\bf s}}}}
\left[
\tilde \alpha^+_{\bf s}({\bm \sigma}) \tilde \phi^+_{\bf n}({\bf q})
+
\tilde \alpha^-_{\bf s}({\bm \sigma}) \tilde \phi^-_{\bf n}({\bf q})
 \right] \\
 \displaystyle \rule{0cm}{0.6cm}
\frac{1}{\sqrt{\tilde \chi^-_{{\bf n},{\bf s}}}}
\left[ \tilde \alpha^+_{\bf s}({\bm \sigma}) \tilde \phi^-_{\bf n}({\bf q})
+
\tilde\alpha^-_{\bf s}({\bm \sigma}) \tilde\phi^+_{\bf n}({\bf q}) \right] ,
\end{array}\right.
\end{equation}
The merits or otherwise of this approximation are discussed
at the end of \S \ref{Sec:exact-vs-approx} below.
Although the left hand side is normalized
by the overall factor of $\surd \tilde\chi^\pm_{{\bf n},{\bf s}}$,
the individual factors on the right hand side are not normalized.
This is essential to the correct formulation of the ansatz
and is emphasized by the tilde.
The un-normalized symmetrized basis functions are
\begin{equation}
\tilde \phi_{\bf n}^\pm({\bf q})
\equiv
\frac{1}{\sqrt{ N!}}
\sum_{\hat{\mathrm P}} (\pm 1)^{p}
\phi_{{\bf n}}(\hat{\mathrm P}{\bf q}),
\end{equation}
and
\begin{equation}
\tilde \alpha_{\bf s}^\pm({\bm \sigma})
\equiv
\frac{1}{\sqrt{ N!}}
\sum_{\hat{\mathrm P}} (\pm 1)^{p}
\alpha_{{\bf s}}(\hat{\mathrm P}{\bm \sigma}) .
\end{equation}
The $\alpha_{{\bf s}}({\bm \sigma})$ and the $\phi_{{\bf n}}({\bf q})$
are normalized.
The $\surd N!$ here is an immaterial constant
that is convenient but not essential.
Respective symmetrization factors for use below
may be defined in terms of these,
\begin{equation}
\chi^\pm_{\bf n}
\equiv
\langle \tilde \phi_{\bf n}^\pm | \tilde \phi_{\bf n}^\pm \rangle
, \mbox{ and }
\chi^\pm_{\bf s}
\equiv
\langle \tilde \alpha_{\bf s}^\pm | \tilde \alpha_{\bf s}^\pm \rangle .
\end{equation}
It is essential  that these symmetrization factors
are not used to normalize the  $\tilde \phi_{\bf n}^\pm({\bf q})$
and the $\tilde \alpha_{\bf s}^\pm({\bm \sigma})$ individually.
The reason for this is that without individual normalization,
all terms in the approximation for
$\Phi_{{\bf n},{\bf s}}^\pm({\bf x})$
have equal weight when written as permutation sums.
If the individual factors were normalized,
then the symmetric factors would have a different weight
to the antisymmetric factors,
and since different products have two, one, or zero of each of these,
individual terms in the permutation sum would have different weights.

This point also explains why the two products
in each line of the approximation are simply added together symmetrically.
In principle, the two products could be superposed
with a relative phase factor
and with a relative probability factor.
The reason they aren't is that ultimately the expression is meant
to approximate a permutation sum in which all terms have equal weight
(apart from the $(-1)^{p}$ for fermions).
These two points will be taken up in the next subsubsection,
and with the concrete example of two particles following that.

The approximation is \emph{sufficient} to ensure
the symmetrization of the spin-position basis functions,
\begin{equation}
\Phi_{{\bf n},{\bf s}}^\pm(\hat{\mathrm P}{\bf x})
=
(\pm 1)^{p} \Phi_{{\bf n},{\bf s}}^\pm({\bf x}),
\end{equation}
as can be confirmed by inspection.

The overall normalization factor
for the symmetrized basis function $\Phi_{{\bf n},{\bf s}}^\pm({\bf x})$
for the approximation is
\begin{eqnarray}
\tilde \chi^\pm_{{\bf n},{\bf s}}
& \equiv &
\left\{ \begin{array} {l}
\langle \tilde \alpha_{\bf s}^+ | \tilde \alpha^+_{\bf s} \rangle \,
\langle \tilde \phi_{\bf n}^+ | \tilde \phi^+_{\bf n} \rangle
+
\langle \tilde\alpha_{\bf s}^- | \tilde \alpha^-_{\bf s} \rangle \,
\langle \tilde\phi_{\bf n}^- | \tilde \phi^-_{\bf n} \rangle
\\  \rule{0cm}{0.45cm}
\langle \tilde \alpha_{\bf s}^+ | \tilde \alpha^+_{\bf s} \rangle \,
\langle \tilde \phi_{\bf n}^- | \tilde \phi^-_{\bf n} \rangle
+
\langle \tilde \alpha_{\bf s}^- | \tilde \alpha^-_{\bf s} \rangle \,
\langle \tilde \phi_{\bf n}^+ | \tilde \phi^+_{\bf n} \rangle
\end{array} \right.
\nonumber \\ & = &
\left\{ \begin{array} {l}
\chi^+_{\bf s} \chi^+_{\bf n} + \chi^-_{\bf s} \chi^-_{\bf n}
\\ \rule{0cm}{0.45cm}
\chi^+_{\bf s} \chi^-_{\bf n} + \chi^-_{\bf s} \chi^+_{\bf n} .
\end{array} \right.
\end{eqnarray}
It should be noted that in certain states the Fermi exclusion principle
means that
$\tilde \alpha_{\bf s}^-({\bm \sigma})$
or $\tilde \phi_{\bf n}^-({\bf q})$ vanish.
In such states $\chi^-_{\bf s}$ and  $\chi^-_{\bf n}$
also respectively vanish.

\subsection{Comparison of Exact and Approximate Forms}
\label{Sec:exact-vs-approx}

One can label the $N!$ permutations of the dynamical variables
${\bm \sigma}_P$ and ${\bf q}_P$, $P = 1,2, \ldots, N!$,
in such a way that the permutation has the same parity as its label.
The exact result is
\begin{equation}
\Phi^\pm_{{\bf n},{\bf s}}({\bf x})
=
\frac{1}{ \sqrt{ N!\chi^\pm_{{\bf n},{\bf s}} } }
\sum_{P=1}^{N!}
(\pm 1)^{P}
\alpha_{{\bf s}}({\bm \sigma}_P)
\phi_{{\bf n}}({\bf q}_P) .
\end{equation}

The approximate expression may be written
\begin{eqnarray}
\Phi_{{\bf n},{\bf s}}^\pm({\bf x})
& = &
\frac{1}{\sqrt{\tilde \chi^\pm_{{\bf n},{\bf s}}}}
\left[
\tilde \alpha^+_{\bf s}({\bm \sigma}) \tilde \phi^\pm_{\bf n}({\bf q})
+
\tilde \alpha^-_{\bf s}({\bm \sigma}) \tilde \phi^\mp_{\bf n}({\bf q})
\right]
\nonumber \\ & = &
\frac{1}{N! \sqrt{\tilde\chi^\pm_{{\bf n},{\bf s}}}}
\sum_{P',P''=1}^{N!} \left[ (\pm 1)^{P''} + (-1)^{P'} (\mp 1)^{P''} \right]
\nonumber \\ && \mbox{ } \times
\alpha_{\bf s}({\bm \sigma}_{P'})  \phi_{\bf n}({\bf q}_{P''})
\nonumber \\ & = &
\frac{2}{N!\sqrt{\tilde\chi^\pm_{{\bf n},{\bf s}}}}
\sum_{P=1}^{N!}
(\pm 1)^{P} \alpha_{{\bf s}}({\bm \sigma}_P) \phi_{{\bf n}}({\bf q}_P)
\nonumber \\ && \mbox{ }
+
\frac{1}{N!\sqrt{\tilde\chi^\pm_{{\bf n},{\bf s}}}}
\sum_{P',P''} \!^{(P' \ne P'')}\,
(\pm 1)^{P''}
\nonumber \\ && \mbox{ } \times
\left[ 1 + (-1)^{P''+P'}  \right]
\alpha_{\bf s}({\bm \sigma}_{P'})  \phi_{\bf n}({\bf q}_{P''}) .
\end{eqnarray}
In the final equality,
the single sum is over terms
where the permutation of the spins is the same
as that of the positions.
This
is the same as the exact result.
(Evidently, if this were the only contribution,
then the normalization constants would be related as
 $ \tilde\chi^\pm_{{\bf n},{\bf s}}
= 4 \chi^\pm_{{\bf n},{\bf s}} /N!$.)
The double sum, where the permutation of the spins differs
from that of the particles,
does not appear in the exact result and is unphysical
in the sense that it disassociates the spin and position of each particle.
It may be noted that only permutations
of spin and position with the same parity
make a non-zero contribution to this double sum.
In the case of $N=2$ no such terms exist,
so the double sum is zero, and the approximation is exact in this case
(see also next).

Notice that the
agreement of part of the approximation with the exact formulation
depends upon using the un-normalized symmetrized basis functions,
$\tilde \alpha^\pm_{\bf s}({\bm \sigma})$
and $\tilde \phi^\pm_{\bf n}({\bf q})$,
and upon superposing the two terms without
any phase factor or probability weight.
As mentioned, the justification for this
is that this procedure weights all terms in the permutation sum equally.

In the light of this analysis of the symmetrized basis functions,
it is worth discussing whether or not the
approximation Eq.~(\ref{Eq:Phi-approx})
has any advantages over the exact form, Eq.~(\ref{Eq:Phi-exact}).
One could argue that the sum of products in the approximate form
has a transparent interpretation that lends itself
to the physical interpretation of symmetrization effects,
and of physical phenomena such as Bose-Einstein condensation
and Fermi exclusion.
Indeed, the long-standing electronic orbital theory for optical spectra
is predicated on this sum of products form,
(which is exact for $N=2$; see also next).
It could also be argued that there could be computational advantages
to obtaining and storing the symmetrized spinless functions,
and at a later stage combining them with the symmetrized spin functions
for different values of $S$.
Finally, it might be argued that exploring the properties
of the  symmetrized spinless functions
leads directly to an understanding of the spatial (or momentum)
localization of symmetrization effects,
which is often missed in the formal treatment of symmetrization.
Such localization shows that in many important terrestrial cases
the nearest neighbor dimers
give the dominant contribution,
and for such pairs the approximation is exact.

\subsection{Simple Example, $N=2$}

For two particles, $N=2$,
the comparison of the exact formulation for symmetrization
with the approximate form can be performed rather directly.
With ${\bf x}' \equiv \hat{\mathrm P}_{12} {\bf x}$,
the exact result is
\begin{equation}
\Phi^\pm_{{\bf n},{\bf s}}({\bf x})
=
\frac{1}{ \sqrt{ 2\chi^\pm_{{\bf n},{\bf s}} } }
\left[
\alpha_{{\bf s}}({\bm \sigma})
\phi_{{\bf n}}({\bf q})
\pm
\alpha_{\bf s}({\bm \sigma}')
\phi_{{\bf n}}({\bf q}')
\right]  .
\end{equation}
The approximation gives
\begin{eqnarray}
\Phi_{{\bf n},{\bf s}}^\pm({\bf x})
& = &
\frac{1}{\sqrt{\tilde \chi^\pm_{{\bf n},{\bf s}}}}
\left[
\tilde \alpha^+_{\bf s}({\bm \sigma}) \tilde \phi^\pm_{\bf n}({\bf q})
+
\tilde \alpha^-_{\bf s}({\bm \sigma}) \tilde \phi^\mp_{\bf n}({\bf q})
\right]
\nonumber \\ & = &
\frac{1}{2\sqrt{\tilde\chi^\pm_{{\bf n},{\bf s}}}}
\left[
\left\{ \alpha_{\bf s}({\bm \sigma}) + \alpha_{\bf s}({\bm \sigma}') \right\}
\left\{ \phi_{\bf n}({\bf q}) \pm \phi_{\bf n}({\bf q}') \right\}
\right. \nonumber \\ && \mbox{ } \left.
+
\left\{ \alpha_{\bf s}({\bm \sigma}) - \alpha_{\bf s}({\bm \sigma}') \right\}
\left\{ \phi_{\bf n}({\bf q}) \mp \phi_{\bf n}({\bf q}') \right\}
\right]
\nonumber \\ & = &
\frac{1}{\sqrt{\tilde\chi^\pm_{{\bf n},{\bf s}}}}
\left[
\alpha_{{\bf s}}({\bm \sigma})
\phi_{{\bf n}}({\bf q})
\pm
\alpha_{\bf s}({\bm \sigma}')
\phi_{{\bf n}}({\bf q}')
\right] .
\end{eqnarray}
This is the same as the exact form.
(Evidently in this case $ \tilde\chi^\pm_{{\bf n},{\bf s}}
= 2 \chi^\pm_{{\bf n},{\bf s}}$.)

One can illustrate the approximation further
by making direct contact with conventional electronic orbital theory
for the simple case of two fermions,
$N=2$, $S=1/2$.
In this case
the symmetrized, un-normalized spin basis functions are
\begin{eqnarray}
\tilde \alpha_{\bf s}^\pm({\bm \sigma})
& \equiv &
\frac{1}{\surd{2}}
\left\{
\alpha_{s_1}(\sigma_1) \alpha_{s_2}(\sigma_2)
\pm
\alpha_{s_1}(\sigma_2) \alpha_{s_2}(\sigma_1)
\right\}
\nonumber \\ & = &
\frac{1}{\surd{2}}
\left\{
\delta_{s_1,\sigma_1}  \delta_{s_2,\sigma_2}
\pm
 \delta_{s_1,\sigma_2}  \delta_{s_2,\sigma_1}
\right\} ,
\end{eqnarray}
and
\begin{equation}
\chi^\pm_{\bf s}
=
\sum_{\hat{\mathrm P}} (\pm 1)^{p}
\langle \hat{\mathrm P} {\bf s} | {\bf s} \rangle
=
1 \pm  \delta_{s_1,s_2} .
\end{equation}
This may be re-written to show the same and different state occupancies
explicitly,
\begin{eqnarray}
\tilde \alpha_{\bf s}^\pm({\bm \sigma})
& = &
\frac{\delta_{s_1,s_2}}{\surd 2}
\left[ \delta_{{\bf s},{\bm \sigma}}
\pm \delta_{{\bf s},{\bm \sigma}} \right]
\nonumber\\ && \mbox{ }
+
\frac{\overline \delta_{s_1,s_2}}{\surd 2}
\left[ \overline \delta_{\sigma_1,\sigma_2}
\left( \delta_{s_1,\sigma_1}
\pm \delta_{s_1, \sigma_2} \right)  \right]
\nonumber \\ & = &
\frac{\delta_{s_1,s_2}}{\surd 2}
\left\{ \begin{array} {l}
2 \delta_{{\bf s},{\bm \sigma}}
\\
0
\end{array} \right.
+
\frac{\overline \delta_{s_1,s_2}}{\surd 2}
\overline \delta_{\sigma_1,\sigma_2}
\left[ \delta_{s_1,\sigma_1}
\pm \overline \delta_{s_1, \sigma_1} \right]
\nonumber \\ & = &
\frac{\delta_{s_1,s_2}}{\surd 2}
\left\{ \begin{array} {l}
2 \delta_{{\bf s},{\bm \sigma}}
\\
0
\end{array} \right.
\nonumber\\ && \mbox{ }
+
\frac{\overline \delta_{s_1,s_2}}{\surd 2}
\left\{ \begin{array} {l}
\overline \delta_{\sigma_1,\sigma_2}
\\
\overline \delta_{\sigma_1,\sigma_2}
\left[ \delta_{s_1,\sigma_1}
- \overline \delta_{s_1, \sigma_1} \right].
\end{array} \right.
\end{eqnarray}
Here and throughout, the complementary Kronecker delta is
$\overline \delta_{j,k} \equiv 1 - \delta_{j,k}$.

For the basis functions in position
one can consider single particle states,
$\phi_{\bf n}({\bf q}) =
\phi_{{\bf n}_1}({\bf q}_1) \phi_{{\bf n}_2}({\bf q}_2)$.
In the case of electronic orbitals,
typically ${\bf n}_j = \{ n_j , l_j, m_j\}$.
Then
\begin{equation}
\chi^\pm_{\bf n}
=
\sum_{\hat{\mathrm P}} (\pm 1)^{p}
\langle \hat{\mathrm P} {\bf n} | {\bf n} \rangle
=
1 \pm \delta_{{\bf n}_1,{\bf n}_2}  ,
\end{equation}
and
\begin{eqnarray}
\tilde \phi^\pm_{\bf n}({\bf q})
& = &
\frac{1}{\surd{2}}
\left[
\phi_{{\bf n}_1}({\bf q}_1) \phi_{{\bf n}_2}({\bf q}_2)
\pm
\phi_{{\bf n}_1}({\bf q}_2) \phi_{{\bf n}_2}({\bf q}_1)
\right]
\nonumber \\ & = &
\frac{\delta_{{\bf n}_1,{\bf n}_2}}{\surd 2}
\left\{ \begin{array} {l}
2 \phi_{{\bf n}_1}({\bf q}_1) \phi_{{\bf n}_2}({\bf q}_2) \\
0
\end{array} \right.
\\ && \mbox{ }\nonumber
+
\frac{\overline \delta_{{\bf n}_1,{\bf n}_2} }{\surd 2}
 \left[
\phi_{{\bf n}_1}({\bf q}_1) \phi_{{\bf n}_2}({\bf q}_2)
\pm
\phi_{{\bf n}_1}({\bf q}_2) \phi_{{\bf n}_2}({\bf q}_1)
\right] .
\end{eqnarray}
For use shortly, it follows that
\begin{eqnarray}
\tilde\phi^+_{\bf n}({\bf q}) \pm \tilde\phi^-_{\bf n}({\bf q})
& = &
\delta_{{\bf n}_1,{\bf n}_2} \sqrt{2}
\phi_{{\bf n}_1}({\bf q}_1) \phi_{{\bf n}_2}({\bf q}_2)
\\ && \mbox{ }\nonumber
+
\overline \delta_{{\bf n}_1,{\bf n}_2} \sqrt{2}
\left\{ \begin{array} {l}
\phi_{{\bf n}_1}({\bf q}_1) \phi_{{\bf n}_2}({\bf q}_2)
\\
\phi_{{\bf n}_1}({\bf q}_2) \phi_{{\bf n}_2}({\bf q}_1) .
\end{array} \right.
\end{eqnarray}

The overall normalization factor is
\begin{eqnarray}
\lefteqn{
\tilde \chi^\pm_{{\bf n},{\bf s}}
} \nonumber \\
& = &
\left\{ \begin{array} {l}
(1 + \delta_{s_1,s_2} )(1 + \delta_{{\bf n}_1,{\bf n}_2} )
+ (1 - \delta_{s_1,s_2} )(1 - \delta_{{\bf n}_1,{\bf n}_2} )
\\ \rule{0cm}{0.45cm}
(1 + \delta_{s_1,s_2} )(1 - \delta_{{\bf n}_1,{\bf n}_2} )
+
(1 - \delta_{s_1,s_2} )(1 + \delta_{{\bf n}_1,{\bf n}_2} )
\end{array} \right.
\nonumber \\ & = &
2 (1 \pm \delta_{s_1,s_2} \delta_{{\bf n}_1,{\bf n}_2} ) .
\end{eqnarray}
For fermions, this vanishes if both are in the same state.
Since the normalization factor appears in the denominator
as the square root,
the symmetrized spin-position basis function,
$\Phi^-_{{\bf n},{\bf s}}({\bf x})$, also vanishes in this case.

Putting these together,
the anti-symmetrized two fermion wave function is
\begin{eqnarray}
\lefteqn{
\psi^-_{{\bf n},{\bf s}}({\bf q},{\bm \sigma})
} \nonumber \\
& = &
\frac{1}{\sqrt{\tilde \chi^-_{{\bf n},{\bf s}} }}
\left[ \alpha^+_{\bf s}({\bm \sigma}) \tilde\phi^-_{\bf n}({\bf q})
+ \alpha^-_{\bf s}({\bm \sigma}) \tilde\phi^+_{\bf n}({\bf q}) \right]
\nonumber \\ & = &
\frac{\surd 2}{\sqrt{\tilde \chi^-_{{\bf n},{\bf s}} }}
\delta_{s_1,s_2}
 \delta_{{\bf s},{\bm \sigma}} \tilde\phi^-_{\bf n}({\bf q})
\nonumber \\ && \mbox{ }
+
\frac{ \overline \delta_{s_1,s_2} }{ \sqrt{ \tilde\chi^-_{{\bf n},{\bf s}} } }
 \frac{1}{\sqrt{2}}
\left[
\delta_{s_1,\sigma_1}  \delta_{s_2,\sigma_2}
\left\{ \tilde\phi^-_{\bf n}({\bf q}) + \tilde\phi^+_{\bf n}({\bf q}) \right\}
\right. \nonumber \\ && \left. \mbox{ }
+  \delta_{s_1,\sigma_2}  \delta_{s_2,\sigma_1}
\left\{ \tilde\phi^-_{\bf n}({\bf q}) - \tilde\phi^+_{\bf n}({\bf q}) \right\}
 \right]
 \nonumber \\ & = &
\frac{\surd 2}{\sqrt{ \tilde \chi^-_{{\bf n},{\bf s}} }}
\delta_{s_1,s_2}
 \delta_{{\bf s},{\bm \sigma}} \tilde\phi^-_{\bf n}({\bf q})
\nonumber \\ && \mbox{ }
+
\frac{ \overline \delta_{s_1,s_2} }{\sqrt{ 2\tilde\chi^-_{{\bf n},{\bf s}} }}
\delta_{s_1,\sigma_1}  \delta_{s_2,\sigma_2}
\left\{
\delta_{{\bf n}_1,{\bf n}_2} \surd 2
\phi_{{\bf n}_1}({\bf q}_1) \phi_{{\bf n}_2}({\bf q}_2)
\right. \nonumber \\ && \left. \mbox{ }
+
\overline \delta_{{\bf n}_1,{\bf n}_2}  \surd 2
\phi_{{\bf n}_1}({\bf q}_1) \phi_{{\bf n}_2}({\bf q}_2)
\right\}
\nonumber \\ &  & \mbox{ }
-
\frac{ \overline \delta_{s_1,s_2} }{\sqrt{ 2\tilde\chi^-_{{\bf n},{\bf s}} }}
\delta_{s_1,\sigma_2}  \delta_{s_2,\sigma_1}
\left\{
\delta_{{\bf n}_1,{\bf n}_2} \surd 2
\phi_{{\bf n}_1}({\bf q}_1) \phi_{{\bf n}_2}({\bf q}_2)
\right. \nonumber \\ && \left. \mbox{ }
+
\overline \delta_{{\bf n}_1,{\bf n}_2}  \surd 2
\phi_{{\bf n}_1}({\bf q}_2) \phi_{{\bf n}_2}({\bf q}_1)
\right\}
 \nonumber \\ & = &
\delta_{s_1,s_2} \delta_{{\bf s},{\bm \sigma}}
\frac{\overline \delta_{{\bf n}_1,{\bf n}_2} }{\surd 2}
 \left[
\phi_{{\bf n}_1}({\bf q}_1) \phi_{{\bf n}_2}({\bf q}_2)
-
\phi_{{\bf n}_1}({\bf q}_2) \phi_{{\bf n}_2}({\bf q}_1)
\right]
\nonumber \\ && \mbox{ }
+
\frac{ \overline \delta_{s_1,s_2} }{ \sqrt{ 2 } }
\delta_{s_1,\sigma_1}  \delta_{s_2,\sigma_2}
\phi_{{\bf n}_1}({\bf q}_1) \phi_{{\bf n}_2}({\bf q}_2)
\nonumber \\ &  & \mbox{ }
-
\frac{ \overline \delta_{s_1,s_2} }{ \sqrt{ 2 } }
\delta_{s_1,\sigma_2}  \delta_{s_2,\sigma_1}
\left\{
\delta_{{\bf n}_1,{\bf n}_2}
\phi_{{\bf n}_1}({\bf q}_1) \phi_{{\bf n}_2}({\bf q}_2)
\right. \nonumber \\ && \left. \mbox{ }
+
\overline \delta_{{\bf n}_1,{\bf n}_2}
\phi_{{\bf n}_1}({\bf q}_2) \phi_{{\bf n}_2}({\bf q}_1)
\right\}.
\end{eqnarray}

For $s_1=s_2$ this is
\begin{eqnarray}
\psi^-_{{\bf n},{\bf s}}({\bf q},{\bm \sigma})
& = &
\frac{1}{\sqrt{2}}
\delta_{{\bf s},{\bm \sigma}}
\overline \delta_{{\bf n}_1,{\bf n}_2}
\left\{
\phi_{{\bf n}_1}({\bf q}_1) \phi_{{\bf n}_2}({\bf q}_2)
\right. \nonumber \\ && \left. \mbox{ }
-
\phi_{{\bf n}_1}({\bf q}_2) \phi_{{\bf n}_2}({\bf q}_1)
\right\}
. 
\end{eqnarray}
This vanishes if ${\bf n}_1 = {\bf n}_2$,
which is just the Fermi exclusion principle.

For ${\bf n}_1 = {\bf n}_2$
one  has
\begin{eqnarray}
\psi^-_{{\bf n},{\bf s}}({\bf q},{\bm \sigma})
& = &
\frac{ \overline \delta_{s_1,s_2} }{\surd 2}
\phi_{{\bf n}_1}({\bf q}_1) \phi_{{\bf n}_2}({\bf q}_2)
\nonumber \\ &&  \mbox{ } \times
\left[
\delta_{s_1,\sigma_1}  \delta_{s_2,\sigma_2}
- \delta_{s_1,\sigma_2}  \delta_{s_2,\sigma_1}
\right]
. 
\end{eqnarray}
This vanishes if $s_1 = s_2$,
which is again the Fermi exclusion principle.
The term in brackets
is equivalent to the so-called singlet state.

For ${\bf n}_1 \ne {\bf n}_2$
one has
\begin{eqnarray}
\psi^-_{{\bf n},{\bf s}}({\bf q},{\bm \sigma})
& = &
\delta_{s_1,s_2} \delta_{{\bf s},{\bm \sigma}} \tilde\phi^-_{\bf n}({\bf q})
 \\ &  & \mbox{ }
+
\frac{ \overline \delta_{s_1,s_2} }{\surd 2}
\delta_{s_1,\sigma_1}  \delta_{s_2,\sigma_2}
\phi_{{\bf n}_1}({\bf q}_1) \phi_{{\bf n}_2}({\bf q}_2)
\nonumber \\ &  & \mbox{ }
-
\frac{ \overline \delta_{s_1,s_2} }{\surd 2}
\delta_{s_1,\sigma_2}  \delta_{s_2,\sigma_1}
\phi_{{\bf n}_1}({\bf q}_2) \phi_{{\bf n}_2}({\bf q}_1)
.\nonumber
\end{eqnarray}
These terms are
equivalent to the so-called triplet state
(the first term represents the $++$ and $--$ states,
and the second and third terms
are a superposition of the $+-$ and $-+$ state).

\end{document}